\documentclass[fleqn,twoside]{article}
\usepackage{espcrc2}

\usepackage{epsfig}
\newcommand{\beq}{\begin{equation}}
\newcommand{\eeq}{\end{equation}}
\newcommand{\bea}{\begin{eqnarray*}}
\newcommand{\eea}{\end{eqnarray*}}
\newcommand{\beqa}{\begin{eqnarray}}
\newcommand{\eeqa}{\end{eqnarray}}

\newcommand{\AmS}{{\protect\the\textfont2
  A\kern-.1667em\lower.5ex\hbox{M}\kern-.125emS}}

\title{Improved Wilson QCD simulations at light quark masses}

\author{A. C. Irving\address[MCSD]{
Theoretical Physics Division,
         Department of Mathematical Sciences,
         University of Liverpool,
         PO Box 147, Liverpool L69 3BX, UK.},
        {\em UKQCD Collaboration}\thanks{supported by PPARC grant
        PP/G/S/1998/00777}}
\begin{document}

\begin{abstract}
We present preliminary results from UKQCD simulations at light quark
masses using two flavours of non-pertubatively improved Wilson fermions.
We report on the performance of the standard HMC algorithm at these
quark masses where $m_\pi/m_\rho<0.5$ in comparison
with simulations using improved staggered quarks. 
\vspace{1pc}
\end{abstract}

\maketitle

\section{SIMULATION PERFORMANCE}
Since 1997, UKQCD
has been conducting dynamical
fermion simulations on a Cray T3E machine using typically
$128$ processors ($35$ Gflops sustained). This has permitted
production studies on lattices up to $16^3\times 32$ and $m_\pi/m_\rho$
down to around 0.60~\cite{csw202}. 
Study groups within the collaboration are now
planning physics programmes for the advent of the QCDOC
multi-Teraflops machine which
is expected to be our main resource from 2003 onwards.
Actions and algorithmic performance are key concerns.

\subsection{HMC performance}
The next figure shows estimates of the work (in Teraflops)
required to obtain a \lq new\rq{} configuration using three of the
main methods currently available.

% 1st Figure
\centerline{\epsfig{file=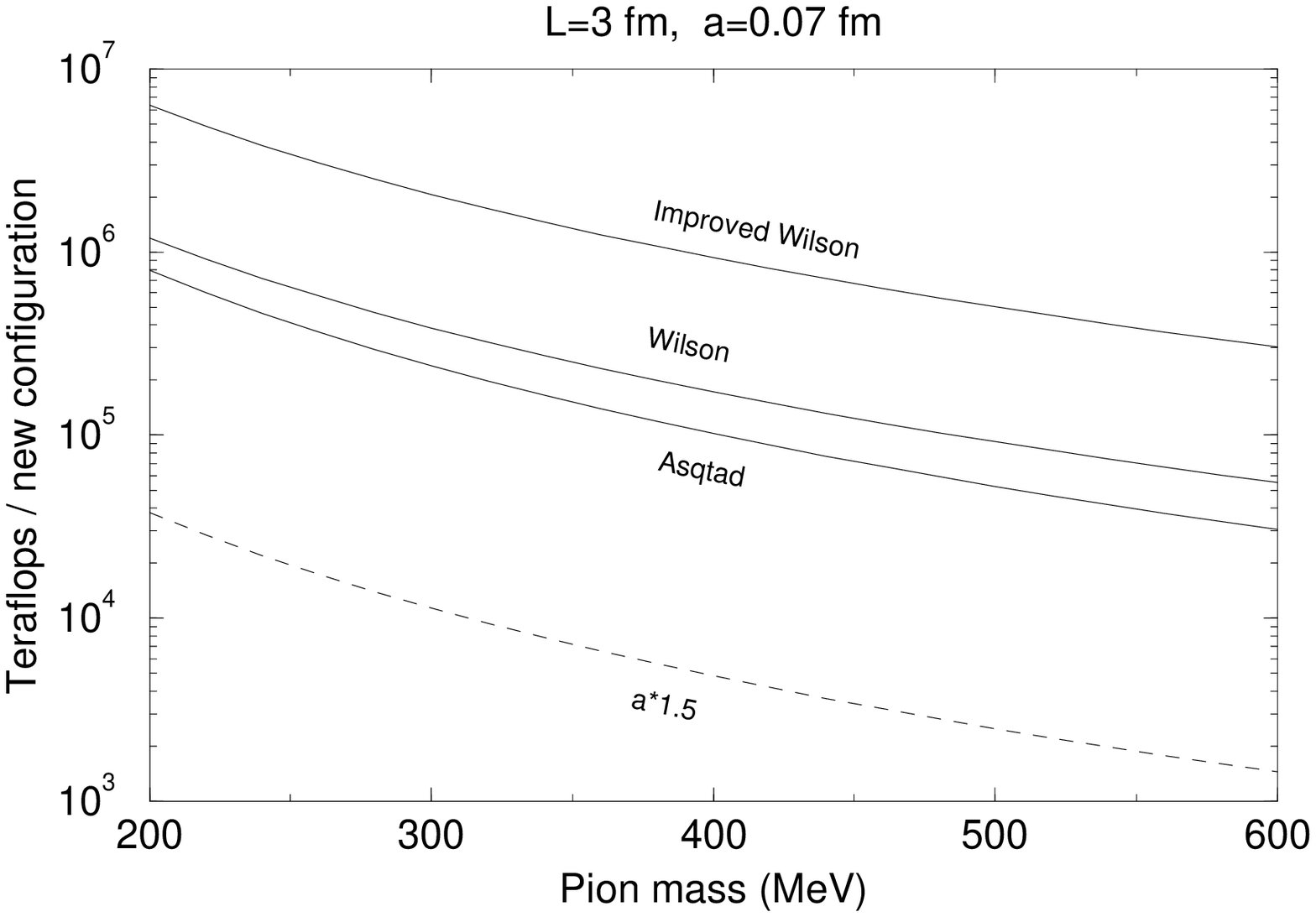,width=240pt}}

The data for the plot was mostly taken from that provided last year
at LAT01 for the session on algorithm performance~\cite{Panel01}.
The top curve corresponds to the
UKQCD \lq clover-Wilson\rq{} estimate:
\beq
{\hbox{Gflops}\over \hbox{config}}=0.157
\Bigl({L\over a}\Bigr)^{3.41}
\Bigl({T\over a }\Bigr)^{1.14}
\Bigl({1\over {am_{\pi}}}\Bigr)^{2.77}\, .
\label{eq:ukqcd}
\eeq
The accompanying curves correspond to 
Wilson fermions as measured by the SESAM collaboration~\cite{SESAM}
and the Asqtad MILC code (\lq order $a^2$ tadpole\rq{}
improved staggered fermions
using the R-algorithm~\cite{MILC01}).
The dashed curve corresponds to the Asqtad data
but where the lattice spacing has been increased by $50\%$.
The Asqtad scaling
curve was taken from the performance data in~\cite{SG02}.
We performed an independent check by carrying out
Asqtad simulations with 2 flavours
on the same size lattice and machine used to produce the UKQCD data.
For this comparison,
a \lq new configuration\rq{}
amounts to 20 MD trajectories
for the Wilson and clover HMC runs and 6 for the Asqtad runs.
These numbers are based on using at least twice the measured integrated
plaquette autocorrelation time in each case.
The motivation for the extra comparison of Wilson
at lattice spacing $a$
with Asqtad at spacing $1.5a$ is the observation
that the improvement for the latter
is much greater (${\cal O} (a^2\alpha_s$)) than for Wilson at a
given spacing $a$. Since the work scales like $~a^{-8}$ this contributes
very significantly to the saving in work at fixed physical lattice size
and $m_\pi$.

The Wilson/SESAM simulation~\cite{SESAM} appears to be $8$ times faster
than the improved Wilson/UKQCD one.
Detailed comparison shows that
a factor of $3.4$ comes from solver iterations (better
preconditioning and solver algorithm and better
underlying condition number);
a factor of $1.8$ from larger MD step size (higher
acceptance, physics);
a factor of $1.3$ from the numerical overhead of the clover term.
These naive comparisons 
do not take into account other factors
such as relative scaling behaviour and accessibility and efficiency
of measurement operators.

\subsection{HMC stability}

These HMC runs are expected to be susceptible
to instabilities (occasional large $\Delta H$ values and zero
acceptance)
when the fermion force term gets too large~\cite{BJoo}. 
We have observed this effect
directly on the clover-Wilson run 
at $\kappa=0.1358$ where
we find it necessary to use a step size of $1/400$. When the
step size is this, or smaller, we find that $64$ bit
for our field storage and matrix-vector manipulations
is required so as to avoid a dramatic
loss of acceptance due to rounding errors. This is so even though we
always use full 64-bit arithmetic and sophisticated summing techniques
in the global summing of our $\Delta H$ (energy difference)
calculations.

\subsection{Run parameters}

The main parameters of the new light quark run are shown
in the following table. 
In physical units, $L\approx 1.5 fm$ and $m_\pi\approx 430$ Mev
before extrapolation.
\begin{center}
\begin{tabular}{|l|l|}
\hline
$(\beta,\kappa)$ & $(5.2,.1358)$\\
trajectories/hour & $3$\\
no. trajectories & $2440$\\
no. configs & $122$\\
$\tau^{\hbox{int}}_{AC}$(plaquette) & $6.9(6)$\\
\hline
$r_0/a$ & $5.3(5)$ \\
$a$ (using $r_0=0.49$ fm) & $0.0925(9)$ fm\\
$am_{PS}$ & $0.207(5)$\\
$m_{PS}/m_V$ & $0.43(2)$\\
$m_{PS} L$ & $3.3(1)$\\
\hline
\end{tabular}
\end{center}
The integrated autocorrelation time for the plaquette
follows the previously observed trend that it
{\em decreases} with decreasing quark mass.
For comparison, $\tau^{\hbox{int}}_{AC}$=$16(3)$
at $\kappa=.1350$ where $m_{PS}/m_V\approx 0.7$.
Although not expected, this trend
can be accommodated in simple models~\cite{csw202}.

The ratio \lq $m_\pi/m_\rho$\rq{} is below the
threshold for the continuum decay $\rho\rightarrow\pi\pi$
but not for decay on this size of lattice.
We are using maximum entropy methods~\cite{MaxEnt} to look for evidence
of decay of the lattice state at the relevant energy.

\section{MEASUREMENTS}

\subsection{Hadron masses and finite size effects}

Since $m_\pi L =3.3(1)$ (significantly below $5$) we anticipate finite
size effects.
For small and intermediate size boxes there is evidence~\cite{Fvol}
that hadron mass corrections follow a power law ($L^{-3}$) rather than
exponential ($\exp{(-L/L_0})$) size dependence.
A simple phenomenological model for this might be
\beq
{{\delta m_h}\over{m_h}}  = {{c_h}\over{(m_\pi L)^3}}\, .
\label{eq:FScor}
\eeq
The dimensionless
parameter $c_h$ is presumably
related to the form factor of the hadron $h$.

The figure below shows a compilation 
(from UKQCD and QCDSF~\cite{QCDSF}) 
of pseudoscalar and vector masses 
(fixed $\beta=5.2$ and $\kappa_s=\kappa_v$). 
The filled points are the raw masses extracted from
multi-channel fits.

% 3rd Figure
\centerline{\epsfig{file=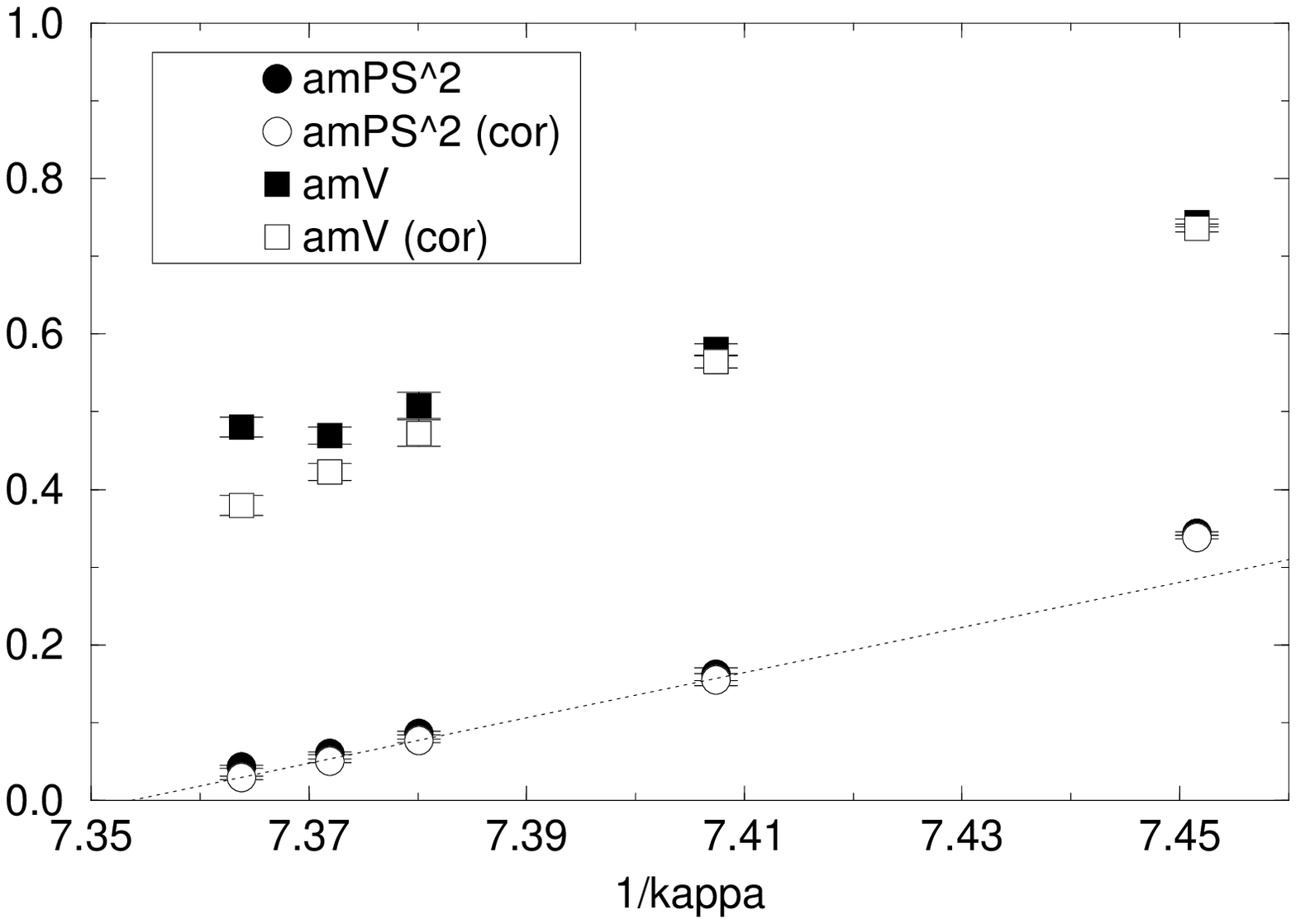,width=240pt}}
Aside from the point at very heavy quark mass ($\kappa=0.1342$),
the $m_{PS}$ data (filled circles) are well fitted by
lowest order chiral PT (linear in $1/\kappa$).

We have estimated $c_\pi\approx 6$ and $c_\rho\approx 7$ from the
finite size (FS) effects reported by the JLQCD Collaboration,
who used the same action (clover-Wilson)~\cite{JLQCD}
on a variety of lattice sizes.
We have checked that the
corrections implied by these values are consistent with the
FS effects which UKQCD has previously measured using
a slightly different action~\cite{csw176}. 
Using this model
to \lq correct\rq{} the measured values, we get the
open points shown in the figure.

The FS-corrected
pseudoscalar points are still well fitted by a straight line
and yield $\kappa_{crit} = 0.135978(9)(55)$.
The first error is statistical and the second
systematic - estimated by including the
heaviest quark point in a quadratic fit.
Taking the above FS corrections
for the pseudoscalar and vector into account
we estimate $m_{PS}/m_V \approx 0.45$
for the lightest quark simulation.

\subsection{Static potential}

The next figure shows the difference of the measured static
potential from the universal bosonic string model
for all UKQCD data sets at fixed $\beta=5.2$~\cite{csw202}
including this latest run (stars)
at $\kappa=0.1358$
and a QCDSF set at $\kappa=0.1340$~\cite{QCDSF}.

% 2nd Figure
\centerline{\epsfig{file=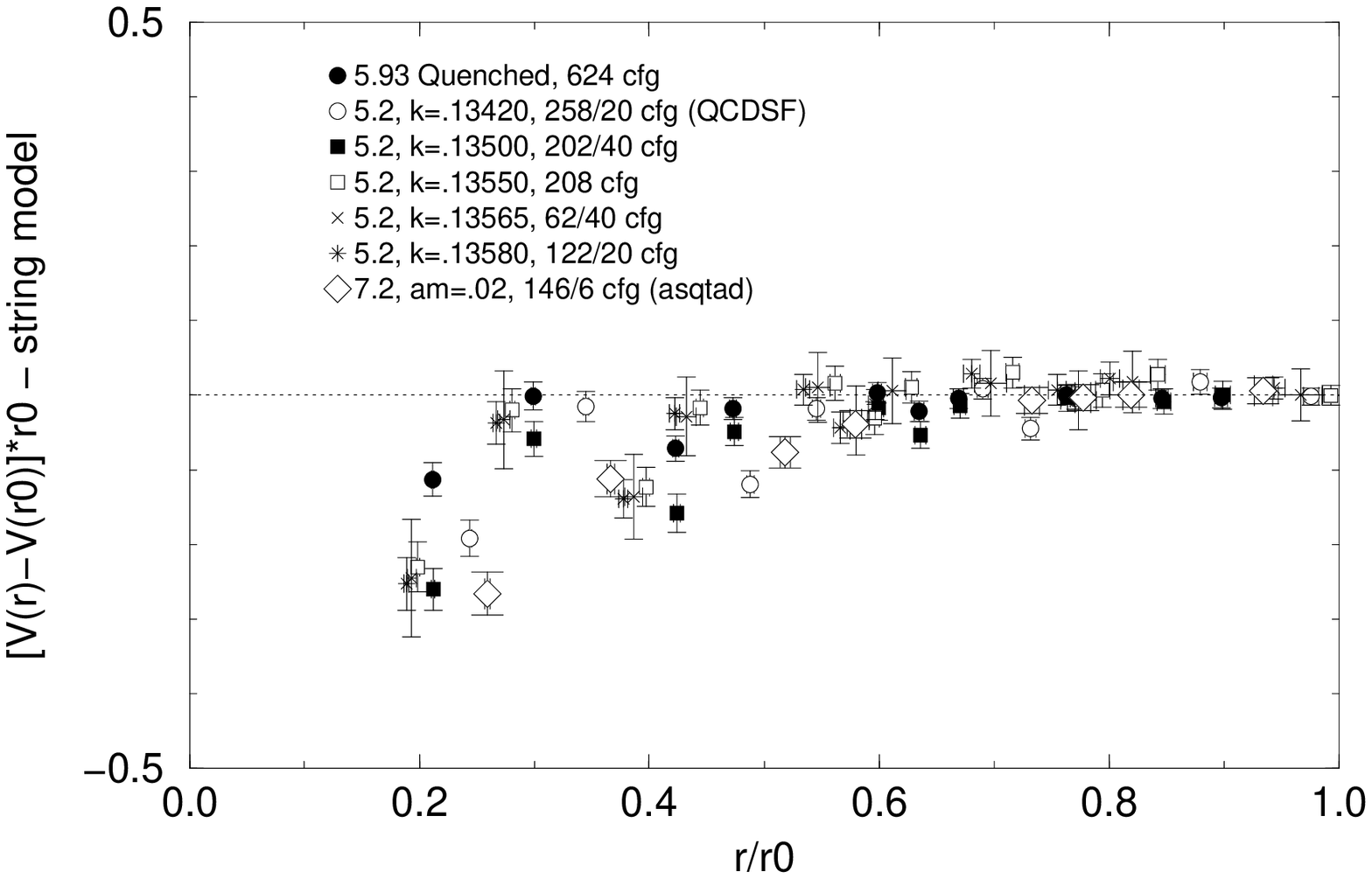,width=240pt}}
Also shown is the corresponding result from a modest run we have made
using the MILC Collaboration's
code at $a\approx 0.130$ fm~\cite{MILC01}.
Violations of rotational symmetry are marked for $r < 0.7 r_0\approx
0.35$
fm for all these coarse lattices.
Although the improved staggered fermion measurements (Asqtad)
are at a coarser lattice spacing, they show smaller
violations of rotational symmetry
than the clover-Wilson data. This is presumably due to the
significantly improved gluonic action.
The Asqtad data (at $m_{PS}/m_V\approx 0.50$) show a greater effect of
the running coupling in the Coulomb term at short distances than
the corresponding clover-Wilson data (lower at short distances).
Within the fixed $\beta$ clover-Wilson set
there is no very clear trend with decreasing quark mass.
The corresponding plot for clover-Wilson data at 
fixed lattice spacing~\cite{csw202} shows more clearly the effect
of the running coupling.

\section{CONCLUSIONS}

We have analysed data from more than 2200 trajectories of
a clover-improved Wilson simulation at $m_q/m_s\approx 0.3$.
We have estimated finite volume corrections to meson masses.
Light quark masses are accessible using clover-improved Wilson fermions
and standard HMC but simulations are slow and can suffer from instabilities.
Improved staggered fermions appear to offer a more efficient
route to configurations with light dynamical quarks, but
physics measurements are in general more complicated.

\section*{Acknowledgements}
I thank my UKQCD colleagues in the dynamical fermions
project, particularly Derek Hepburn, Craig McNeile and Steve Miller.
I thank Thomas Lippert and Steve Gottlieb for detailed insights
into their algorithm performance figures,
and the MILC Collaboration for the use of
their simulation code.

% references

\end{document}